\documentclass[%
preprint,%
aps,%
prl,%
showpacs,%
floatfix,%
superscriptaddress,%
nofootinbib,%
preprintnumbers,%
showkeys
]{revtex4}

\usepackage[dvips]{graphicx}
\usepackage{color}
\usepackage{epsfig}
\usepackage{amsmath}
\usepackage{amssymb}
\usepackage{booktabs}
\usepackage{array}
\usepackage{afterpage}

\preprint{STUPP-08-195}
\preprint{FTUV-08/2205}
\preprint{IFIC/08-41}



\begin{document}

\title{ %
  Big-bang nucleosynthesis and the relic abundance of dark matter
  in a stau-neutralino coannihilation scenario
}

\author{Toshifumi Jittoh}
\email{jittoh@krishna.th.phy.saitama-u.ac.jp}
\affiliation{Department of Physics, Saitama University, 
        Shimo-Okubo, Sakura-ku, Saitama, 338-8570, Japan}
\author{Kazunori Kohri}
\email{k.kohri@lancaster.ac.uk}
\affiliation{Physics Department, Lancaster University LA1 4YB, UK}
\author{Masafumi Koike}
\email{koike@krishna.th.phy.saitama-u.ac.jp}
\affiliation{Department of Physics, Saitama University, 
        Shimo-Okubo, Sakura-ku, Saitama, 338-8570, Japan}
\author{Joe Sato}
\email{joe@phy.saitama-u.ac.jp}
\affiliation{Department of Physics, Saitama University, 
        Shimo-Okubo, Sakura-ku, Saitama, 338-8570, Japan}
\author{Takashi Shimomura}
\email{takashi.shimomura@uv.es}
\affiliation{
Departament de F\'{i}sica Te\`{o}rica and IFIC, Universitat de Val\`{e}ncia-CSIC,
E-46100 Burjassot, Val\`{e}ncia, Spain
}
\author{Masato Yamanaka}
\email{masa@krishna.th.phy.saitama-u.ac.jp}
\affiliation{Department of Physics, Saitama University, 
        Shimo-Okubo, Sakura-ku, Saitama, 338-8570, Japan}

\begin{abstract}
  A scenario of the big-bang nucleosynthesis is analyzed within the
  minimal supersymmetric standard model, which is consistent with a
  stau-neutralino coannihilation scenario to explain the relic
  abundance of dark matter.
  We find that we can account for the possible discrepancy of the
  abundance of $\mathrm{^{7}Li}$ between the observation and the
  prediction of the big-bang nucleosynthesis by taking the mass of the
  neutralino as $300 \, \mathrm{GeV}$ and the mass difference
  between the stau and the neutralino as $(100 \, \textrm{--} \,
  120) \, \mathrm{MeV}$.
  We can therefore simultaneously explain the abundance of the dark
  matter and that of $\mathrm{^{7}Li}$ by these values of parameters.
  The lifetime of staus in this scenario is predicted to be $O(100 \,
  \textrm{--} \, 1000) \, \mathrm{sec}$.
\end{abstract}

\maketitle

The supersymmetric models are attractive candidates of the theory
beyond the standard model.
While no experiments so far have found any evidence of the supersymmetry
, the Large Hadron Collider is expected to find its first
signal in the near future.
On the other hand, the analysis of the cosmological implications of
the supersymmetry is an approach complementary to the direct search.
The lightest supersymmetric particle (LSP) is stabilized by the
$R$ parity and naturally qualifies as the cosmological dark matter.
A possible candidate of the LSP, and hence of the dark matter, is the
neutralino $\tilde{\chi}^{0}$.
The neutralinos as the LSP with their mass of $O(100 \, \mathrm{GeV})$
can be responsible for the present abundance of the dark matter when
the mass of the next-lightest supersymmetric particle (NLSP) is close
to that of the LSP's and allow them to coannihilate with each other in
the early universe.

We put this coannihilation scenario into the perspective of the
big-bang nucleosynthesis (BBN).
The recent results from the Wilkinson Microwave Anisotropy Probe
 experiment~\cite{Dunkley:2008ie}, combined with the standard
BBN scenario, suggest twice or thrice 
as much abundance of $\mathrm{^{7}Li}$ as
suggested from the observation of metal-poor halo
stars~\cite{Li7obs,Bonifacio:2006au,7LiProblem-other}.
This discrepancy may imply that the $\mathrm{^{7}Li}$ nuclei were
destructed in the BBN era through processes of physics beyond the
standard model although there  might still be a possible astrophysical process
to deplete $^{7}$Li  uniformly~\cite{Korn:2006tv}.
We introduced in Ref.~\cite{Jittoh:2007fr} a scenario in which an
exotic negatively-charged massive particle form a bound state
with a nucleus and therethrough initiate the destruction.
There we analyzed the Minimal Supersymmetric Standard Model 
with the coannihilation scenario where the NLSP is the stau
$\tilde{\tau}$.
Staus serve as charged massive particles that trigger the destruction of
$\mathrm{^{7}Li}$ through the interaction
\begin{equation}
\begin{split}
  \mathcal{L}_{\textrm{int}}
  & =
  \tilde{\tau}^{\ast} \overline{\tilde{\chi}^{0}}
  (g_{\textrm{L}} P_{\textrm{L}} + g_{\textrm{R}} P_{\textrm{R}})
  \tau
  +
  \frac{4 G_{\textrm{F}}}{\sqrt{2}}
  \nu_{\tau} \gamma^{\mu} P_{\textrm{L}}\tau J^{\textrm{had}}_{\mu}
  \\ &
  +
  \frac{4 G_{\textrm{F}}}{\sqrt{2}}
  (\bar{l} \gamma^{\mu} P_{\textrm{L}} \nu_{l})
  (\bar{\nu}_{\tau} \gamma_{\mu} P_{\textrm{L}} \tau)
  + \textrm{H.c.},
\end{split}
\label{eq:stau-interaction}
\end{equation}
where
$G_{\textrm{F}} = 1.166 \times 10^{-5} \, \mathrm{GeV^{-2}}$ is the
Fermi constant,
$P_{\textrm{L}}$ and $P_{\textrm{R}}$ are the chiral projection
operators,
$l \in \{ \textrm{e}, \mu \}$,
$g_{\textrm{L}}$ and $g_{\textrm{R}}$ are the coupling constants,
and $J^{\textrm{had}}_{\mu}$ is the hadron current.
The stau $\tilde{\tau}$ in Eq.~(\ref{eq:stau-interaction}) is the mass
eigenstate, which is given by the linear combination of the
superpartner of the left-handed tau $\tilde{\tau}_{\textrm{L}}$ and
that of the right-handed tau $\tilde{\tau}_{\textrm{R}}$ as
\begin{math}
  \tilde{\tau}
  =
    \tilde{\tau}_{\textrm{L}} \cos \theta_{\tau}
  + \tilde{\tau}_{\textrm{R}}
    \mathrm{e}^{-\mathrm{i} \gamma_{\tau}} \sin \theta_{\tau},
\end{math}
where $\theta_{\tau}$ and $\gamma_{\tau}$ are the left-right mixing
angle and CP-violating phase, respectively.
The formation of a stau-$\mathrm{^{7}Be}$ bound state
\begin{math}
  \tilde{\tau} + \mathrm{^{7}Be}
  \to
  (\tilde{\tau} \, \mathrm{^{7}Be}) + \gamma
\end{math}
is immediately followed by an internal conversion process
\begin{math}
  (\tilde{\tau} \, \mathrm{^{7}Be})
  \to
  \tilde{\chi}^{0} + \nu_{\tau} + \mathrm{^{7}Li},
\end{math}
and subsequent spallation of $\mathrm{^{7}Li}$ by the energetic
protons in the background.
We assumed in Ref.~\cite{Jittoh:2007fr} the rapid formation of the
stau-nucleus bound state and ignored the effect of the expansion of
the Universe, as the use of the Saha equation implies.

In the present paper, we improve the previous analysis by considering the expansion effect of the Universe.
The Boltzmann equation is employed instead of the Saha equation to
estimate the stau-nucleus bound states.
We also include the resonant formation of the bound state pointed out
in Ref.~\cite{Bird:2007ge}.
We thereby show that the LSP and NLSP%
, both with mass of $O(100 \, \mathrm{GeV})$,
can account for the problem of the dark matter and that of the
abundance of $\mathrm{^{7}Li}$.
The relevant parameters in considering our BBN scenario are the mass
difference between staus and neutralinos $\delta m \equiv
m_{\tilde{\tau}} - m_{\tilde{\chi}^{0}}$, where $m_{\tilde{\tau}}$ and
$m_{\tilde{\chi}^{0}}$ are the masses of staus and neutralinos,
respectively, and the yield value of the staus at the {\it freeze-out} time
\begin{math}
  Y_{\tilde{\tau}, \textrm{FO}}
  \equiv
  {n_{\tilde{\tau}} / s} \, \vert_{\textrm{Freeze Out}},
\end{math}
where $n_{\tilde{\tau}}$ and $s$ are the densities of the number of
staus and the entropy, respectively.
Other parameters are fixed throughout this paper to $\theta_{\tau} =
\pi/3$, $\gamma_{\tau} = 0$, and $m_{\tilde{\chi}^{0}} = 300 \,
\mathrm{GeV}$.
By varying the values of $\delta m$ and $Y_{\tilde{\tau},
  \textrm{FO}}$, we search for the parameter region that can account
for the present abundance of $^{7}\mathrm{Li}$.
Staus play a major role in the BBN when $\delta m$ is small so that
staus become longevous enough to survive until the BBN era.
The lifetime of staus indeed becomes $100 \, \mathrm{sec}$ or longer
when $\delta m \lesssim 100 \, \mathrm{MeV}$~%
\cite{Jittoh:2005pq,Jittoh:2007fr}.

\begin{figure*}
\includegraphics[width=0.95\textwidth]{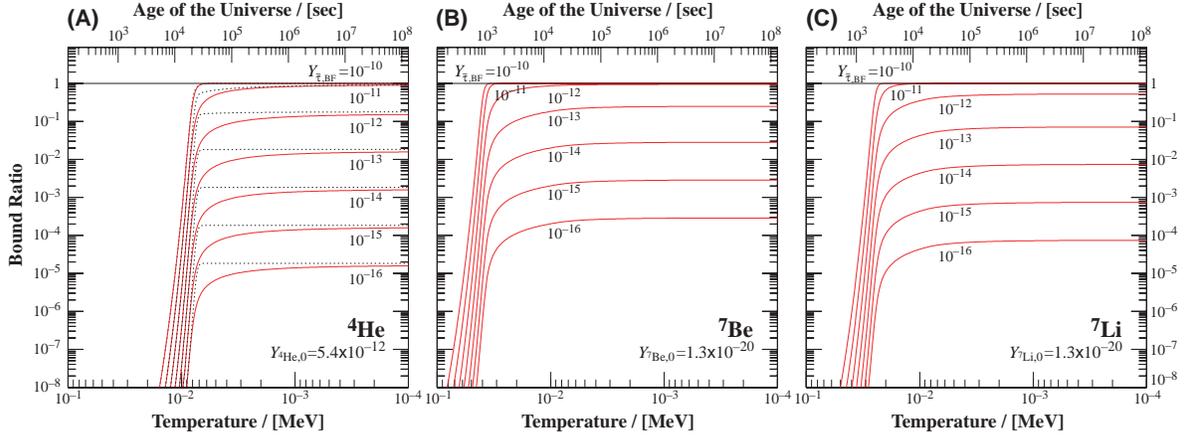}
\caption{\label{fig:bound-ratio}%
  Evolutions of the bound ratio of the nuclei $\mathrm{^{4}He}$,
  $\mathrm{^{7}Be}$, and $\mathrm{^{7}Li}$.
  We vary the abundance of the stau at the time of the formation of
  the bound state from $10^{-10}$ to $10^{-16}$ in each figure.
  In Fig.~\ref{fig:bound-ratio}(a), we also plotted by dotted lines corresponding
  curves predicted using the Saha equation for reference.}
\end{figure*}

We show in Fig.~\ref{fig:bound-ratio} the evolutions of the bound
ratios of $\mathrm{^{4}He}$, $\mathrm{^{7}Li}$, and $\mathrm{^{7}Be}$,
where we define the bound ratio by the number density of a nucleus
that forms a bound state with a stau, divided by the total number
density of that nucleus.
We trace the evolution of the number density of the stau-nucleus bound
states by the Boltzmann equation, using the cross sections shown in
Ref.~\cite{Kohri:2006cn}.

The yield value of staus at the time of \textit{the formation of the
  bound state} with nuclei $t_{\textrm{BF}}$, which we denote by
$Y_{\tilde{\tau}, \textrm{BF}}$, is changed from $10^{-10}$ to
$10^{-16}$ in each figure.
It is related with $Y_{\tilde{\tau},\textrm{FO}}$ using the lifetime
of stau $\tau_{\tilde{\tau}}$ as
\begin{equation}
  Y_{\tilde{\tau}, \textrm{BF}}
  =
  Y_{\tilde{\tau}, \textrm{FO}} \,
  \mathrm{e}^{-t_{\textrm{BF}} / \tau_{\tilde{\tau}}}.
\end{equation}

The bound ratio of $\mathrm{^{4}He}$ shown in
Fig.~\ref{fig:bound-ratio}(a) is crucial to estimate the creation
rate of $\mathrm{^{6}Li}$ due to the catalyzed fusion
process~\cite{Pospelov:2006sc,Hamaguchi:2007mp}
\begin{math}
  (\tilde{\tau} \, \mathrm{^{4}He}) + \mathrm{D}
  \to
  \mathrm{^{6}Li} + \tilde{\tau},
\end{math}
while that of $\mathrm{^{7}Be}$ shown in Fig.~\ref{fig:bound-ratio}(b) is necessary to evaluate
the reduction rate of the $\mathrm{^{7}Li}$.
Since the present $\mathrm{^{7}Li}$ originates from the primordial
$\mathrm{^{7}Be}$, the abundance of $\mathrm{^{7}Li}$ is reduced as follows.
We first convert $\mathrm{^{7}Be}$ into $\mathrm{^{7}Li}$ by an
internal conversion process 
\begin{math}
  (\tilde{\tau} \, \mathrm{^{7}Be})
  \to
  \tilde{\chi}^{0} + \nu_{\tau} + \mathrm{^{7}Li},
\end{math}
and successively destruct the daughter $\mathrm{^{7}Li}$ by either a
collision with a background proton or a subsequent internal conversion
\begin{math}
  (\tilde{\tau} \, \mathrm{^{7}Li})
  \to
  \tilde{\chi}^{0} + \nu_{\tau} + \mathrm{^{7}He}.
\end{math}
The $\mathrm{^{7}Be}$ is efficiently reduced if its bound ratio
plotted in Fig.~\ref{fig:bound-ratio}(b) is of $O(1)$.
The successive destruction of $\mathrm{^{7}Li}$ by internal conversion
is effective when its bound ratio plotted in Fig.~\ref{fig:bound-ratio}(c) is also of
$O(1)$.
We find in Figs.~\ref{fig:bound-ratio}(b) and \ref{fig:bound-ratio}(c) that both bound ratios of
$\mathrm{^{7}Be}$ and $\mathrm{^{7}Li}$ are of $O(1)$ when
\begin{math}
  Y_{\tilde{\tau}, \mathrm{FO}} \gtrsim (10^{-13} \, \textrm{--} \, 10^{-12}).
\end{math}

\begin{figure}[h]
\begin{center}
\includegraphics[width=70mm]{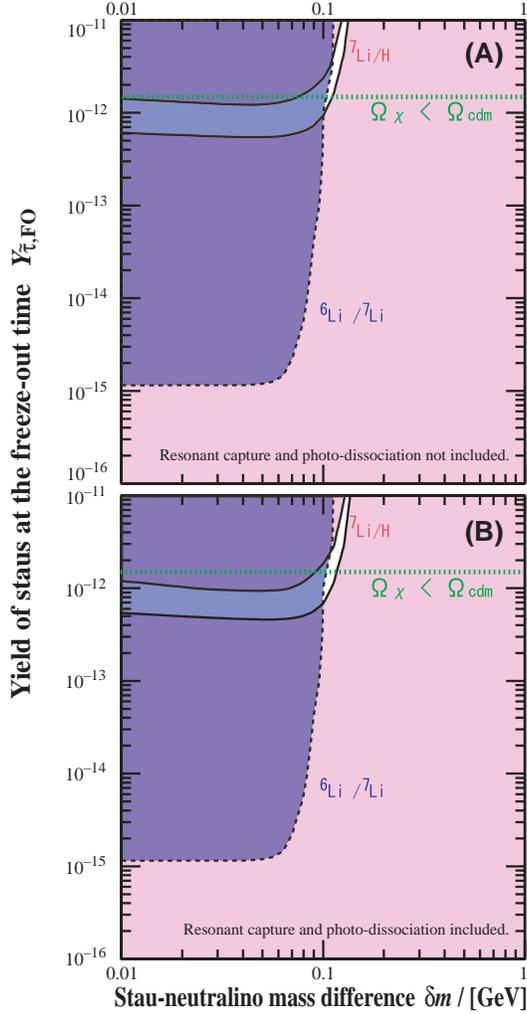}
\caption{%
  Allowed region in $\delta m$-$Y_{\tilde{\tau},\mathrm{FO}}$ plane.
  The resonant capture and the photo dissociation are (a) not
  included, and (b) included.
  The white region is the parameter space, which is consistent with all
  the observational abundance including that of
  $\mathrm{^{7}Li/H}$.
  The region enclosed by dashed lines is excluded by the
  observational  abundance of
  $\mathrm{^{6}Li/^{7}Li}$~\cite{Asplund:2005yt}, and the one enclosed by
  solid lines are allowed by those of
  $\mathrm{^{7}Li/H}$~\cite{Bonifacio:2006au}.
  The thick dotted line is given by the upper bound of the yield value
  of dark matter.
  This line gives the upper bound of $Y_{\tilde{\tau}, \mathrm{FO}}$.
}
\label{y_deltam}
\end{center}
\end{figure}
The parameter region that can solve the $^{7}\mathrm{Li}$ problem is
numerically calculated in the $(\delta m, Y_{\tilde{\tau},
  \textrm{FO}})$ plane and presented in Figure~\ref{y_deltam}, in which
Fig.~\ref{y_deltam}(a) does not include effects of the resonant formation and
photo-dissociation processes of the bound state pointed out in
Ref.~\cite{Bird:2007ge}, while Fig.~\ref{y_deltam}(b) includes these effects.
The white region is the parameter space, which is consistent with all
the observational abundance including that of
$\mathrm{^{7}Li/H}$.
The region enclosed by dashed lines is excluded by the
observational  abundance of
$\mathrm{^{6}Li/^{7}Li}$~\cite{Asplund:2005yt}, and the one enclosed by
solid lines are allowed by those of
$\mathrm{^{7}Li/H}$~\cite{Bonifacio:2006au}.
The thick dotted line is given by the upper bound of the yield value
of dark matter
\begin{equation}
  Y_{\mathrm{DM}} = 4.02 \times 10^{-12} 
  \Bigl( \frac{\Omega_{\mathrm{DM}} h^{2}}{0.110} \Bigr)
  \Bigl( \frac{m_{\mathrm{DM}}}{10^{2} \, \mathrm{GeV}} \Bigr)^{-1},
\label{eq:YDM}
\end{equation}
taking $\Omega_{\rm DM}h^{2} = 0.1099 +  0.0124
$ (upper bound of $95\%$
confidence level)~\cite{Dunkley:2008ie} and $m_{\mathrm{DM}} =
m_{\tilde{\chi}^{0}}$.
This line gives the upper bound of $Y_{\tilde{\tau}, \mathrm{FO}}$,
since the supersymmetric particles after their freeze-out consist of
not only staus but neutralinos as well in our scenario.

  The allowed region shown in Fig.~\ref{y_deltam} lies at $\delta m
  \simeq (100 \, \textrm{--} \, 120) \, \mathrm{MeV}$, which is tiny
  compared with $m_{\tilde{\chi}^{0}} = 300 \, \mathrm{GeV}$.
  These values of parameters allow the coannihilation between
  neutralinos and staus, and thus can account also for the abundance
  of the dark matter.
  We therefore find that the values of $m_{\tilde{\chi}^{0}} = 300 \,
  \mathrm{GeV}$ and $\delta m \simeq 100 \, \mathrm{MeV}$ can
  simultaneously explain the abundance of dark matter and of
  $\mathrm{^{7}Li}$.

We compare the Figs.~\ref{y_deltam}(a) and \ref{y_deltam}(b) to find that the allowed region is
shifted downward in Fig.~\ref{y_deltam}(b).
Of the two processes included in Fig.~\ref{y_deltam}(b), the resonant formation
process makes the bound ratio larger, the value of $Y_{\tilde{\tau}}$
smaller, and push the allowed region downward in the figure.
On the other hand, the photo dissociation process makes the bound
ratio smaller through the destruction of the bound state, makes the value
of $Y_{\tilde{\tau}}$ larger, and push the allowed region upward.
We thus find that the resonant formation of the bound state is
relevant while the photo dissociation is inconsequential.

The qualitative feature of the allowed region is explained from the
following physical consideration.
First, we note that $Y_{\tilde{\tau}, \mathrm{FO}} \gtrsim (10^{-13} \,
\textrm{--} \, 10^{-12})$ is required so that a sufficient number of
bound state $(\tilde{\tau} \, \mathrm{^{7}Be})$ is formed to destruct
$\mathrm{^{7}Be}$ by the internal conversion into $\mathrm{^{7}Li}$.
The daughter $\mathrm{^{7}Li}$ is broken either by an energetic proton
or by the internal conversion
\begin{math}
  (\tilde{\tau} \, \mathrm{^{7}Li})
  \to
  \tilde{\chi}^{0} + \nu_{\tau} + \mathrm{^{7}He},
\end{math}
and consequently $\mathrm{^{7}Li/H}$ is reduced.
Bearing this physical situation in mind, 
we consider parameter regions in detail.  
\begin{enumerate}
\item $\delta m \gtrsim 120 \, \mathrm{MeV}$.\\
Since the staus decay before they form a bound state with
$\mathrm{^{7}Be}$, the value of $Y_{\tilde{\tau}, \mathrm{BF}}$ is
much lower than $10^{-13}$ and hence the abundance of neither
$\mathrm{^{7}Be}$ nor $\mathrm{^{7}Li}$ is reduced.
Therefore this parameter region
 is excluded.
\item $100 \, \mathrm{MeV} \lesssim \delta m \lesssim 120 \,
\mathrm{MeV}$.\\
The staus are just decaying at the formation time of the bound state.
The necessary condition of $Y_{\tilde{\tau}, \mathrm{BF}} \sim
10^{-13}$ can still be retained even in a case where the value of
$Y_{\tilde{\tau}, \mathrm{FO}}$ is sufficiently large.
The allowed region in this area of $\delta m$ thus bends upward.
In this region, a daughter $\mathrm{^{7}Li}$ from the internal
conversion of $(\tilde{\tau} \, \mathrm{^{7}Be})$ is broken mainly by
an energetic proton.
\item $Y_{\tilde{\tau}, \mathrm{FO}} \lesssim 10^{-13}$.\\
In this case $Y_{\tilde{\tau}, \mathrm{BF}}$ is necessarily less than
$10^{-13}$, and the bound ratio of $\mathrm{^{7}Li}$ and
$\mathrm{^{7}Be}$ are much less than $O(1)$ as seen in
Fig.~\ref{fig:bound-ratio}.
Therefore, the final abundance of $\mathrm{^{7}Li}$ is not reduced
sufficiently.
This parameter region is thus excluded.
\item $Y_{\tilde{\tau}, \mathrm{FO}} > 10^{-12}$ and $\delta m < 100 \, \mathrm{MeV}$.\\
In this region $Y_{\tilde{\tau}, \mathrm{BF}} = Y_{\tilde{\tau},
  \mathrm{FO}} > 10^{-12}$ and hence the bound ratio of
$\mathrm{^{7}Be}$ is $1$ (see Fig.~\ref{fig:bound-ratio}).
It means that $\mathrm{^{7}Be}$ and consequently $\mathrm{^{7}Li}$ are
destructed too much.
Hence, the upper-left region is excluded.
\item $\delta m \lesssim 100 \, \mathrm{MeV}$ and $Y_{\tilde{\tau},
  \mathrm{FO}} \gtrsim 10^{-15}$.\\
In this region,
the stau acquires the long lifetime enough to form a bound state
$(\tilde{\tau} \, \mathrm{^{4}He})$.
Then the catalyzed fusion process $(\tilde{\tau} \, \mathrm{^{4}He}) +
\mathrm{D} \rightarrow \mathrm{^{6}Li} + \tilde{\tau}$ leads to the
overproduction of $\mathrm{^{6}Li}$ and to the disagreement to the
observational limit.
Therefore, this parameter region
is excluded, which is consistent with calculations by
Ref.~\cite{KKMstau}.
\end{enumerate}
  Excluding all the parameter regions described above, we obtain a
  small allowed region of $m_{\tilde{\chi}^{0}} \simeq
  m_{\tilde{\tau}} \simeq 300 \, \mathrm{GeV}$ and $\delta m = (100 \,
  \textrm{--} \, 120) \, \mathrm{MeV}$ as presented in
  Fig.~\ref{y_deltam}, and these values are at the same time
  consistent to the coannihilation scenario of the dark matter.

\vspace*{3mm}
  We obtained a strict constraint on the mass of the neutralinos and
  staus by improving %
an analysis of a solution to the overproduction problem of
$\mathrm{^{7}Li} $ or $\mathrm{^{7}Be}$ through the internal
conversion in stau-nucleus bound states,
\begin{math}
  (\tilde{\tau} \, \mathrm{^{7}Be})
  \to
  \tilde{\chi}^{0} + \nu_{\tau} + \mathrm{^{7}Li}
\end{math}
and
\begin{math}
  (\tilde{\tau} \, \mathrm{^{7}Li}) 
  \to
  \tilde{\chi}^{0} + \nu_{\tau} + \mathrm{^{7}He},
\end{math}
given in Ref.~\cite{Jittoh:2007fr}.
We included the resonant capture process of $\mathrm{^{7}Be}$ and
photo-dissociation process pointed out in Ref.~\cite{Bird:2007ge}.
We also took into account the expansion of the Universe by an
explicit use of the Boltzmann equation instead of the Saha equation to obtain
a more accurate number of the stau-nucleus bound states.
By varying the yield value of the stau at its freeze-out time, we
found that most of $\mathrm{^{7}Li}$ and $\mathrm{^{7}Be}$ nuclei form
a bound state with a stau for $Y_{\tilde{\tau}, \mathrm{BF}} \gtrsim
(10^{-12} \, \textrm{--} \, 10^{-13})$.
Taking the values of $m_{\tilde{\chi}^{0}} = 300 \, \mathrm{GeV}$,
$\theta_{\tau} = \pi/3$, $\gamma_{\tau} = 0$, and $\eta = (6.225 \pm
0.170) \times 10^{-10}$~\cite{Dunkley:2008ie}, we compared the
primordial abundances with and without the resonant capture 
and/or  photo dissociation, and found
that the resonant capture process is relevant while the
photo dissociation process of the bound state is inconsequential.
We obtained a parameter region consistent with the observed abundance
of $\mathrm{^{7}Li}$ within 
\begin{math}
  Y_{\tilde{\tau}, \textrm{FO}}
  = (7 \, \textrm{--} \, 10) \times 10^{-13}
\end{math}
and $\delta m =(100 \, \textrm{--} \, 120) \, \mathrm{MeV}$.
The region of $\delta m \le 100 \, \mathrm{MeV}$ is excluded due to
the overproduction of $\mathrm{^{6}Li}$ by the catalyzed fusion.
Furthermore, the parameter region obtained in this paper lies in the
coannihilation region, which can explain the relic abundance of dark
matter.
Therefore, the stau with $m_{\tilde{\tau}}\sim 300 \, \mathrm{GeV}$ and
$\delta m \sim 100 \, \mathrm{MeV}$ can simultaneously solve the
problems on the relic abundance of the light elements and the dark
matter.
As shown in Ref.~\cite{Jittoh:2005pq}, the stau with $m_{\tilde{\tau}} =
300 \, \mathrm{GeV}$ and $\delta m = 100 \, \mathrm{MeV}$ has the
lifetime of $O(100 \, \textrm{--} \, 1000) \, \mathrm{sec}$.
It is very possible that Large Hadron Collider will find some staus with a very long
lifetime~\cite{longstau}.

We need further improvement on our analysis to obtain a more precise
result of the mass and the mass difference.
We have to derive $Y_{\tilde{\tau}}$ as a function of the parameters
in the Lagrangian, although we regarded $Y_{\tilde{\tau}}$ as a free
input parameter in this paper.
Then, we can determine the allowed region of $\delta m$ and
$m_{\tilde{\chi}^{0}}$ more precisely by varying other parameters such
as $\theta_{\tau}$ and $\gamma_{\tau}$.
We leave this for our future work.

\begin{acknowledgments}
  The work of K.~K. was supported in part by PPARC Grant
  No.~PP/D000394/1, EU Grant No.~MRTN-CT-2006-035863, the European
  Union through the Marie Curie Research and Training Network
  ``UniverseNet,'' MRTN-CT-2006-035863.
  The work of T.~J. was financially supported by the Sasakawa
  Scientific Research Grant from The Japan Science Society.
  The work of J.~S. was supported in part by the Grant-in-Aid for the
  Ministry of Education, Culture, Sports, Science, and Technology,
  Government of Japan Contact Nos. 20025001, 20039001, and 20540251.
  The work of T.~S. was supported in part by MEC and FEDER (EC) Grants
  No.~FPA2005-01678.
  The work of M.~Y. was supported in part by the Grant-in-Aid for the
  Ministry of Education, Culture, Sports, Science, and Technology,
  Government of Japan Contact No. 2007555.
\end{acknowledgments}

%%%%%%%%%%%%%%%%%%%%%%%%%%%%%%%%%%%%%%%%%%%%%%%%%%%%%%

\end{document}